\documentclass[reprint, amsmath,amssymb,aps,showkeys]{revtex4-2}

\usepackage[dvipdfmx]  {graphicx}
\usepackage[dvipsnames]{xcolor}  
\usepackage[separate-uncertainty=true]{siunitx}    
\usepackage{graphicx} 
\usepackage{subfig}
\usepackage{dcolumn}  
\usepackage{bm}       
\usepackage{color}    
\usepackage{ulem}     
\usepackage{comment}  

\newcommand{\del}   [1]{\textrm{d}#1}
\newcommand{\derv}  [2]{ \frac{\mathrm{d}#1}{\mathrm{d}#2} }

\newcommand{\subrm}[1]{_{\mathrm{#1}}}
\newcommand{\frf}     {f_{\mathrm{rf}}}

\newcommand{\refv}[1] {#1_{\mathrm{ref}}}

\begin{document}

\preprint{APS/123-QED}

\title {Wide frequency-range acceleration using second harmonic rf buckets in fixed field accelerators}
\author{T.~Uesugi}
\email {uesugi.tomonori.2n@kyoto-u.ac.jp}
\author{Y.~Ishi, Y.~Mori}
\affiliation{Institute for Integrated Radiation and Nuclear Science, Kyoto University\\
             Kumatori-cho, Sennan-gun, Osaka 590-0494 Japan}
 
\date{\today} 

\begin{abstract}
We propose a novel acceleration scheme for fixed-field accelerators (FFAs), in which rf buckets with harmonic numbers $h = 1$ and $h = 2$ are time-sequenced to form a single, continuous acceleration path.  
This approach completes acceleration in two rf frequency sweeps, thereby reducing the total frequency sweep range and shortening the repetition period.  
The feasibility of this method is demonstrated through longitudinal simulations based on parameters of the FFA at the Institute for Integrated Radiation and Nuclear Science, Kyoto University (KURNS).  
We also establish operational conditions under which the second harmonic rf bucket remains stable and practically usable.
\end{abstract}


\maketitle

\section{Introduction \label{sec:Introduction} }

A fixed-field accelerator (FFA) is a type of circular accelerator that employs a static guide field, allowing closed orbits over a wide range of particle energies.  
In FFAs, the closed orbit and revolution frequency are globally defined as functions of particle momentum, in contrast to conventional synchrotrons, where they are defined only in the vicinity of the synchronous momentum at each instant. 
Owing to their large momentum acceptance, FFAs can support a variety of acceleration schemes~\cite{Symonrf}.  

Modern scaling-type FFAs can achieve a momentum acceptance exceeding a factor of three~\cite{WideP,KURNSFFA}, corresponding to a similarly wide acceleration energy range.  
Such a broad energy range naturally requires a wide rf frequency sweep 
to maintain synchronization during acceleration in FFAs where the revolution frequency varies significantly with particle momentum. 
As the required frequency range becomes broader, the operational burden on the rf system increases accordingly.  
Motivated by this, we developed a new acceleration concept that effectively limits the required rf sweep range while maintaining the wide energy acceptance inherent to FFAs.  

Notably, FFAs can even accommodate multiple beams with different central energies, enabling unique acceleration schemes
(e.g.,~\cite{Stacking,MultiFish}).  
One such scheme is multi-beam acceleration, which was demonstrated in Ref.~\cite{MultiFish} using a two-component rf field.  
Beam-signal measurements confirmed that each rf component independently controlled the corresponding beam bunch, enabling simultaneous acceleration of two beams within a single ring, and thereby doubling the effective output repetition rate.  

Building on this concept, the present work introduces a new approach that realizes a similar dual-bucket operation using the second-harmonic rf buckets, which are inherently available in the accelerator.  
In general, particles are accelerated when the rf frequency satisfies the synchronization condition 
$\frf=hf_{\textrm{rev}}$, where $h$ is the harmonic number and $f_{\textrm{rev}}$ is the revolution frequency of a particle.  
Although these buckets correspond to different synchronous energies, their acceleration ranges can be connected, provided that the final energy of one bucket matches the initial energy of the other bucket.  
Accordingly, the required condition is: 
  \begin{equation}
  f\subrm{fin} =
    \begin{cases} 
      \ 2f\subrm{ini}			& \  \text{(below the transition energy)}\,, \\
      \ \frac{1}{2}f\subrm{ini}	& \  \text{(above the transition energy)}\,,
    \end{cases}
  \label{eq:feq2f}  
  \end{equation}
where $f\subrm{ini}$ and $f\subrm{fin}$ denote the initial and final rf frequencies, respectively.  

For instance, below the transition energy, the beam is first accelerated by the second harmonic bucket up to its final energy, then recaptured by the fundamental bucket, and accelerated further in the next rf frequency sweep cycle.  
A smoothing region is required at the connection point, where the rf voltage is gradually reduced (adiabatic debunching) and then increased (adiabatic capture), 
so that the handover between buckets occurs while the beam is effectively coasting, with neither harmonic active.  
As a result, the overall rf sweep range is inherently limited to a factor of two---a favorable outcome in terms of rf system simplification.  

Note that the use of higher harmonics here does not involve waveform shaping via harmonic superposition, as in conventional multi-harmonic rf systems~\cite{BucketShaping,Painting}.
It also differs from harmonic number jump acceleration~\cite{HNJ}, in which the rf frequency is fixed and the harmonic number seen by the particles changes from turn to turn.  
In contrast, the proposed method employs an rf frequency sweep, with the acceleration proceeding successively through rf buckets of different harmonic numbers.  
While the present approach is conceptually related to the harmonic ratcheting scheme~\cite{Ratchet}, it differs in that 
the same rf pattern can simultaneously serve as both $h=1$ and $h=2$, enabling two beams at different energies to be accelerated concurrently.

Our scheme, referred to as \textit{barber-pole acceleration} (BPA), enables beam acceleration from the synchronous energy corresponding to $\frac{1}{2}f\subrm{ini}$ to that corresponding to $2f\subrm{ini}$, using only the upper segment of the rf frequency range, from $f\subrm{ini}$ to $2f\subrm{ini}$.  
This effectively extends the accessible energy range without requiring a wider rf sweep span.  
Furthermore, although each acceleration takes two sweep cycles, beam injection can still occur once every sweep, allowing the beam delivery rate to be maintained.  
Compared to the original multi-beam approach, in which multiple rf frequency components are superposed, 
the proposed scheme simplifies the system and allows the available rf power to be concentrated into a single frequency component.  
As the ratio between the rf and beam revolution frequencies remains exactly 1 or 2, the resonance condition for rf knock-out~\cite{Stacking} is never met, unless the betatron tune itself is an integer.  
Furthermore, when Eq.~(\ref{eq:feq2f}) is satisfied, the synchronous energy ranges of the fourth harmonic naturally connect to those of the second harmonic, and so on recursively for all $2^n$-th harmonics, as far as closed orbits exist.  
Therefore, even if the frequency range corresponding to the beam energy extends beyond a factor of four, the required rf frequency range remains limited to a factor of two.

Despite its advantages, the scheme has inherent limitations due to its lack of flexibility.  
Once the rf amplitude and frequency pattern are fixed, both the fundamental and second harmonic buckets are simultaneously defined, leaving no room for independent optimization.  
This leads to two potential issues: drift of the synchronous phase, and mismatch in longitudinal acceptance between the two buckets.  

Numerical and simulation studies were conducted to quantitatively assess these concerns.  
Transverse dynamics and space-charge effects are neglected to focus on the fundamental longitudinal behavior of the scheme.  
Throughout this study, the scaling law of FFA~\cite{Okawa,Symon,Kolomensky} is adopted to model the revolution frequency as a function of particle momentum.  
This assumption provides a consistent framework to determine the synchronous energies corresponding to given rf frequencies.

The remainder of the paper is organized as follows.  
Section~\ref{sec:Buckets} presents numerical estimates of the synchronous energies, synchronous phases, and bucket areas for the second harmonic rf field, and identifies the accelerator parameter range in which both the fundamental and second harmonic buckets remain operational.  
Section~\ref{sec:Simulation} reports multi-particle simulations using a model accelerator, demonstrating successful acceleration without significant beam loss.  
Section~\ref{sec:Discussion} reviews the results, extends the analysis to higher harmonics and above-transition scenarios, and discusses possible system refinements and beam stability issues under high-intensity conditions.  
Finally, Section~\ref{sec:Conclusion} summarizes the study and outlines the future directions.

\section{Higher harmonic rf-buckets \label{sec:Buckets} }

\subsection{synchronous energies}

\begin{figure}[thbp]
  \centering
  \includegraphics[scale=1.0]{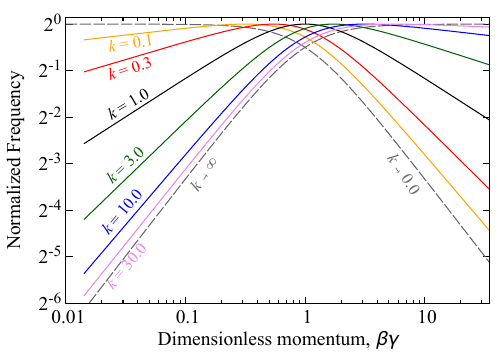}
  \caption{Normalized revolution frequencies in a scaling FFA plotted as a function of $\beta\gamma$.  
  Each curve was normalized so that its maximum or asymptotic value was 1.  
  Different colors indicate different values of the field index $k$.  }
  \label{fig:logFofG}
  \end{figure}

In a scaling FFA, the circumference of the closed orbit for a particle with momentum $p$ is proportional to $p^{\alpha}$, 
where $\alpha=\frac{1}{k+1}$ is the momentum compaction factor.  Here, $k$ is called the field index, a machine-specific constant.  
Under this condition, the revolution frequency scales as 
  \begin{equation}
  f = C\, \frac{\ \ (\beta\gamma)^{1-\alpha}}{E}\,, 
  \label{eq:FofE}
  \end{equation}
where $C$ is a constant, and $\beta$, $\gamma$ and $E$ denote the relative velocity, Lorentz factor, and 
total energy, respectively. 
Figure~\ref{fig:logFofG} illustrates how the functional dependence of $f$ on $\beta\gamma$ varies with the field index $k$. 
The slope of the log-log plot of $f$ vs momentum $p$ corresponds to the negative of the local slippage factor, $\eta$, which is analytically given by $\eta=\alpha-\frac{1}{\gamma^2}$. 

In our scheme, the rf frequency ($\frf$) must be swept over a range spanning a factor of two. 
To ensure successful acceleration, the revolution frequency must be a monotonic function of energy 
within the relevant range, either increasing ($\eta<0$, below transition) or decreasing ($\eta>0$, above transition).   
For a given rf frequency, a set of synchronous energies $E_h$ arises, each satisfying 
  \begin{equation}
  \frf = h f(E_h)\quad=\ h\times C\frac{\ \ (\beta_h\gamma_h)^{1-\alpha}}{E_h}\,, 
  \label{eq:Harmonics}
  \end{equation}
where the subscript $h$ indicates evaluation of the synchronous energy corresponding to the $h$th harmonic.  

The proposed scheme can be particularly useful when an accelerator must operate in an energy region where the revolution frequency is highly sensitive to changes in $\beta\gamma$.  
As shown in Fig.~\ref{fig:logFofG}, such situation typically arises in high-$k$ machines at low momenta, 
and in low-$k$ machines at high momenta.

\subsection{synchronous phases}

Sweeping the rf frequency $\frf$ causes corresponding changes in the synchronous energies $E_h$ for all $h$.
It follows from Eq.~(\ref{eq:Harmonics}) that the synchronous energy increases per turn as 
  \begin{equation*}
  \Delta E_h= \frac{1}{f_h}\derv{E_h}{t} \quad
            =\ \frac{h\beta_h^2E_h}{-\eta_h}\,\frac{\dot{f}_{\mathrm{rf}}}{\frf^2}\,, 
  \end{equation*}
where a dot denotes a time derivative.  
Note that $\Delta E_h$ is related to the $h$-th harmonic synchronous phase $\phi_h$ via $\Delta E_h=q\overline{V}\,\sin\phi_h$, where $q$ and $\overline{V}$ are the particle charge and the rf voltage amplitude, respectively.  

We introduce the acceleration ratio $R_h$, defined by
  \begin{equation}
  R_h = \ \frac{\Delta E_h}{\Delta E_1}\ 
      = h\,\frac{\beta_h^2\gamma_h/\eta_h} {\beta_1^2\gamma_1/\eta_1}\,. 
  \label{eq:DefineRh}
  \end{equation}
Since $\overline{V}$ is common to all harmonics, the synchronous phase for general $h$ can be expressed as 
  \begin{equation}
  \sin\phi_h = R_h\,\sin\phi_1\,.  
  \label{eq:SinPh}
  \end{equation}
If $R_h$ remains approximately constant over the relevant momentum range, it becomes possible to maintain both $\phi_1$ and $\phi_h$ at nearly constant values simultaneously.  
  \begin{figure}[thbp]
  \centering
  \includegraphics[scale=1.0]{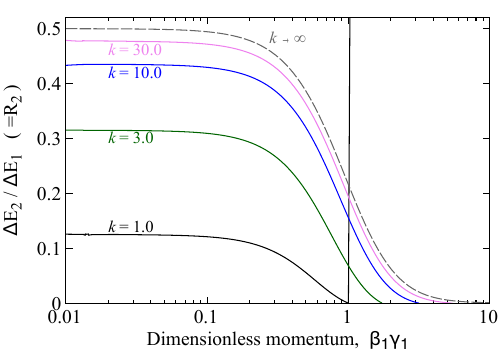}
  \caption{Ratio of one-turn energy gains, $\Delta E_2/\Delta E_1$, 
  as a function of the dimensionless synchronous momentum $\beta\gamma$, in the below-transition regime. 
  Different colors indicate different values of the field index~$k$. 
  }
  \label{fig:PlotR2}
  \end{figure}

Figure~\ref{fig:PlotR2} shows $R_2$ as a function of $\beta_1\gamma_1$ below the transition.  
As seen in the figure, $R_2$ becomes nearly constant in the nonrelativistic limit ($\beta\gamma\ll 1$).  
Although similar trends for general $h$ are not shown explicitly, their asymptotic values can be derived from Eq.~(\ref{eq:Harmonics}) as 
  \begin{equation}
  R_h \simeq h^{-\frac{k+2}{k}} \qquad \text{for } \beta\gamma \ll 1 \,.
  \label{eq:RhApproach}
  \end{equation}
For a discussion of the above-transition case, see Sec.~\ref{sec:Above}.

\subsection{bucket areas and acceptances}

We now examine the longitudinal acceptance at synchronous energies corresponding to higher harmonics.  
Ideally, the acceptance should remain comparable to that at $h=1$ to ensure consistent beam transmission across different harmonic numbers.  
A significant reduction in acceptance at any harmonic would create a bottleneck, ultimately limiting the number of particles that can be accelerated across the full energy range.  

Assuming that the synchronous energy behaves as an adiabatic invariant, 
the rf bucket area---expressed in units of action---is given by Ref.~\cite{SYLee} as 
  \begin{equation}
  A_h 
  = \frac{8\sqrt{2}}{2\pi\frf}\sqrt{\frac{\beta_h^2E_h\,q\overline{V}}{\pi h|\eta_h|}}
    \times a(\phi_h)\,.
  \label{eq:Area}
  \end{equation}
Here, $a(\phi_h)$ is defined by 
  \begin{equation}
  a(\phi_h) = \int\frac{\left\{\cos\phi+\cos\phi_h+\left(\phi+\phi_h-\pi\right)\sin\phi_h\right\}^{\frac{1}{2}}}
                       {4\sqrt{2}}\ 
              \textrm{d}\phi\,, 
  \label{eq:BucketFormFactor}
  \end{equation}
where the integration is performed over the domain containing $\phi_h$ for which square root is real.  
According to Ref.~\cite{SYLee}, the approximation 
  \begin{equation}
  a(\phi) \sim \frac{1-\sin\phi}{1+\sin\phi}
  \label{eq:ApproxA}
  \end{equation}
holds within \qty{2}{\percent} for $0\le\phi\le\frac{\pi}{2}$.  
Since the stable area at energy $E_h$ is repeated $h$ times around the ring, the ratio of longitudinal acceptance is given by 
  \begin{align}
  S_h=\quad
  \frac{hA_h}{A_1}
  &= h\sqrt{\frac{\beta_h^2\gamma_h/h\eta_h}{\beta_1^2\gamma_1/\eta_1}}\,\times
     \frac{a(\phi_h)}{a(\phi_1)}\nonumber\\
  &\simeq \sqrt{R_h}\,\frac{(1-R_h\sin\phi_1)(1+\sin\phi_1)}{(1+R_h\sin\phi_1)(1-\sin\phi_1)}\,.  
  \label{eq:Sh}
  \end{align}

  \begin{figure}[thbp]
  \centering
  \includegraphics[scale=1.0]{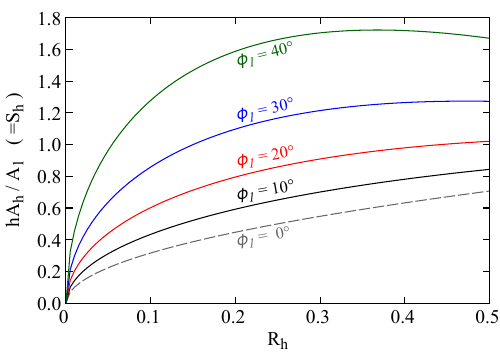}
  \caption{Comparison of longitudinal acceptances between $h=1$ and general $h$, plotted as a function of $R_h$.  }
  \label{fig:RelativeArea}
  \end{figure}

Figure~\ref{fig:RelativeArea} shows this ratio as a function of $R_h$ for various values of $\phi_1$.  
Only the region $R_h<\num{0.5}$ is shown, reflecting the assumption that the system operates in the low-energy regime (below transition).  
At $\phi_1=0$, the ratio $S_h$ reduces to $\sqrt{R_h}$, which remains less than unity throughout the domain considered.   
As $\phi_1$ increases, the difference between $\phi_1$ and $\phi_h$ becomes more pronounced, leading to an increase in $S_h$.  
As seen in the figure, the ratio $S_h$ can be close to unity for a given $R_h$ with an appropriate choice of $\phi_1$; 
for instance, choosing $\phi=\ang{20}$ keeps $S_h$ within approximately $1\pm\qty{10}{\percent}$ in the range $\num{0.3}<R_h<\num{0.5}$.  
For $h=2$, this range corresponds to $k>\num{2.714}$, based on the asymptotic approximation in Eq.~(\ref{eq:RhApproach}).  

The results presented in this section are summarized as follows:
  \begin{itemize}
  \item The BPA scheme is suitable for accelerators that must operate in energy regions 
  where the revolution frequency changes significantly with momentum, as typically observed in machines 
  with a large $k$ at a small $\beta\gamma$ or a small $k$ at large $\beta\gamma$ (see Fig.~\ref{fig:logFofG}).  
  \item The synchronous phases can be held approximately constant 
  if the energy gain ratio $R_2$ remains nearly constant over the operating range (see Fig.~\ref{fig:PlotR2}).
  \item To ensure comparable longitudinal acceptances across harmonics, 
  the value of $\phi_1$ should be appropriately chosen in relation to $R_2$ (see Fig.~\ref{fig:RelativeArea}).
  \end{itemize}

\section{Simulation \label{sec:Simulation} }

\subsection{simulation setup}

Numerical simulations were performed in the longitudinal phase space 
to assess the validity of the proposed scheme more precisely.  
The model machine was based on the existing scaling-type proton FFA at the Institute for Integrated Radiation and Nuclear Science, Kyoto University (KURNS)~\cite{KURNSFFA,Pyeon}.  
The field index is $k = \num{7.645}$, and the final kinetic energy is \qty{100}{\MeV}, corresponding to a revolution frequency of \qty{3.856}{MHz}.  
If this energy is sufficiently low for the asymptotic form in Eq.~(\ref{eq:RhApproach}) to apply, 
the corresponding limit of $R_2$, determined by $k$, is approximately \num{0.4}.  
Substituting this value into Eq.~(\ref{eq:Sh}) yields an acceptance ratio of \num{0.98} for $\phi_1=\ang{20}$.  

In our scheme, the initial rf frequency must be \qty{1.928}{\mega\hertz}, which is exactly half the final frequency.  
While this frequency corresponds to the initial frequency of the $h=1$ bucket, the $h=2$ bucket starts at \qty{0.964}{\mega\hertz}, for which no closed orbit exists in the actual KURNS-FFA.  
Therefore, the frequency--energy relation was extrapolated to the low-energy region under the assumption of perfect scaling.  

  \begin{figure}[thb]
  \centering
  \includegraphics[scale=1.0]{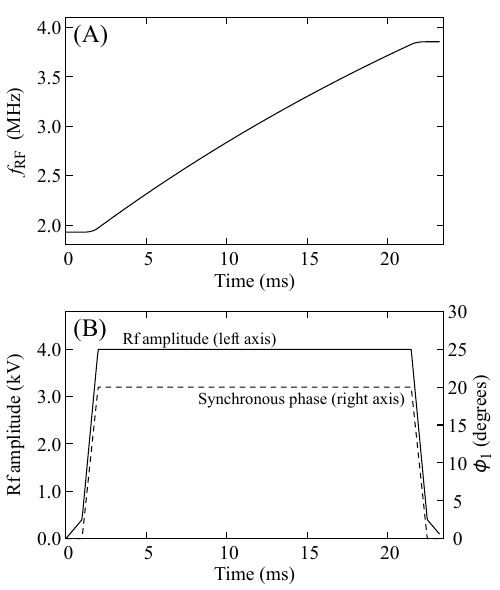}
  \caption{Simulated rf pattern: (A) frequency; (B) rf amplitude (solid line) and synchronous phase (dashed line).   }
  \label{fig:SimrfPattern}
  \end{figure}
  \begin{figure}[thb]
  \centering
  \includegraphics[scale=1.0]{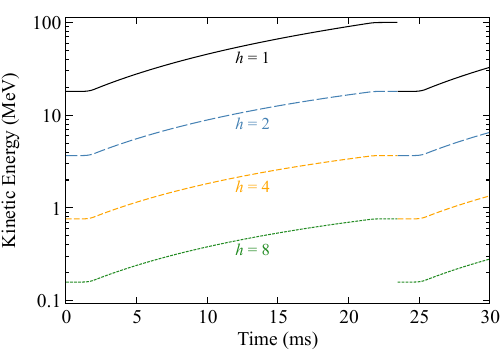}\\
  \caption{Synchronous energies as a function of time for $h = 1$, 2, 4, and 8.  
           Each curve shows the evolution of the synchronous particle's kinetic energy under the rf pattern 
           in Fig.~\ref{fig:SimrfPattern}.  } 
  \label{fig:SimEs8}
  \end{figure}

Figure~\ref{fig:SimrfPattern} illustrates the rf pattern used in the simulation.  
The rf frequency $\frf(t)$ was determined based on the amplitude functions $\overline{V}(t)$ and the synchronous phase $\phi_1(t)$, both shown in Fig.~\ref{fig:SimrfPattern}(B) ( See Appendix~\ref{sec:PatternEdit} ).  
These functions are defined as piecewise linear in time $t$, with parameters listed in Table~\ref{tab:rfParameters}.  
In the main portion of the sweep, the rf amplitude was fixed at \qty{4}{\kilo\volt} and the accelerating phase was held constant at $\phi_1 = \ang{20}$.  
Smoothing segments were added at both ends to mitigate bucket mismatch at the transitions between sections with different $h$.  
(The parameters of the smoothing region were not optimized in the present study; the trade-off between capture efficiency and capture time in adiabatic rf capture is discussed in standard textbooks, e.g., Ref.~\cite{SYLee}. )
The total sweep time of this rf pattern is $T = \qty{23.475}{\milli\second}$, at which point the synchronous energy exactly reaches \qty{100}{\MeV}.  
The rf burst was repeated continuously without any idle intervals.  

The synchronous energies for $h = 1$, $2$, $4$, and $8$ under this rf pattern are shown in Fig.~\ref{fig:SimEs8}, starting from their respective minima: \num{18.065}, \num{3.670}, \num{0.758}, and \qty{0.159}{\MeV}. 
For comparison, if the same acceleration from \num{3.670} to \qty{100}{MeV} were carried out using a conventional method, the rf frequency would need to sweep from \qty{0.964}{\mega\hertz}, resulting in a total acceleration time of \qty{30.67}{\milli\second}.  

  \begin{table}[bth]
  \centering
  \caption{rf amplitude and synchronous phase functions used in the simulation.  
           Both were defined as piecewise linear functions of time $t$ over five intervals. }
  \begin{tabular}{ccc}
  \hline
  Time interval 						& $\overline{V}(t)$							& $\phi_1(t)$\\
  \hline
  $[  \qty{0}{\ms},\   \qty{1}{\ms}]$ 	& $\qty{0  }{\kV}\rightarrow\qty{0.4}{\kV}$	& $\ang{0}$\\
  $[  \qty{1}{\ms},\   \qty{2}{\ms}]$  	& $\qty{0.4}{\kV}\rightarrow\qty{4  }{\kV}$	& $\ang{0}\rightarrow\ang{20}$\\
  $[  \qty{2}{\ms},\ T-\qty{2}{\ms}]$	& $\qty{4  }{\kV}$							& $\ang{20}$\\
  $[T-\qty{2}{\ms},\ T-\qty{1}{\ms}]$	& $\qty{4  }{\kV}\rightarrow\qty{0.4}{\kV}$	& $\ang{20}\rightarrow\ang{0}$\\
  $[T-\qty{1}{\ms},\ T             ]$  	& $\qty{0.4}{\kV}\rightarrow\qty{0  }{\kV}$ & $\ang{0}$\\
  \hline
  \end{tabular}
  \label{tab:rfParameters}
  \end{table}

In the simulation, the evolution of the energy and the ring azimuth of each particle was tracked at every zero crossing of the $\frf$ waveform (See Appendix~\ref{sec:Algorithm}).  
As the initial condition, 10,000 macroparticles were randomly generated using the following distributions: 
The kinetic energy followed a parabolic distribution centered at \qty{3.670}{\MeV}, with a standard deviation of \qty{0.3}{\percent} of that value. 
The azimuth was drawn from a uniform distribution over the interval $[0,2\pi]$.

\subsection{particle tracking results}

\begin{figure}[thb]
  \centering
  \includegraphics[scale=1.0]{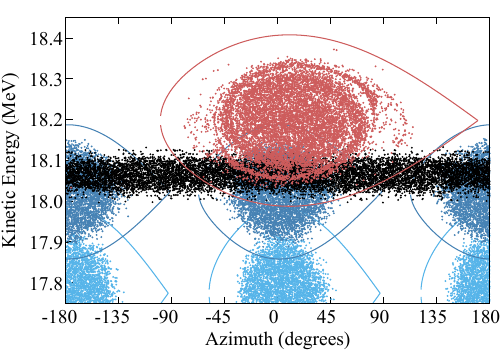}
  \caption{Macroparticle distributions at the boundary energy.  
  Azimuth values originally in [0,\ang{360}], are folded using their periodicity.  
  Colors indicate the time relative to $T$: 
  \num{-2.0}(purple), \num{-1.5}(blue), \num{0.0}(black) and \qty{+1.5}{\ms}(red), respectively, 
  where $T = \qty{23.475}{\milli\second}$.  Curves show the corresponding rf buckets. 
  }
  \label{fig:Transit}
  \end{figure}

The following section presents representative results from the simulation performed under the conditions described in the previous subsection. 
We first examined whether the transition of particles between acceleration stages occurred near the boundary energy, as expected.  
Figure~\ref{fig:Transit} shows the longitudinal phase space around the boundary energy, with the particles plotted at  simulated times close to $T$.  
As illustrated in Fig.\ref{fig:Transit}, the particles are accelerated by the second-harmonic rf buckets, become debunched at the boundary energy, and are subsequently recaptured by the fundamental rf bucket.  

\begin{figure}[thb]
  \centering
  \includegraphics[scale=1.0]{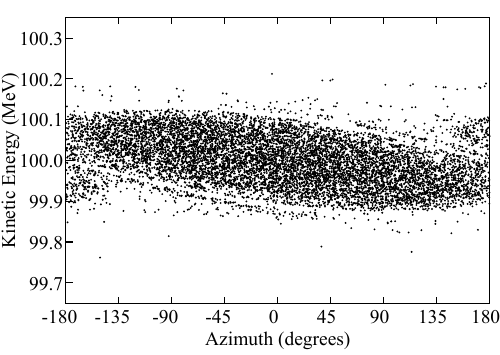}
  \caption{Final macroparticle distribution obtained at simulation time $2T$.  }
  \label{fig:FinalDist}
  \end{figure}

We then evaluated whether the particles reach the final energy.  
After an additional time $T$, more than \qty{99.9}{\percent} of the macroparticles were successfully accelerated 
to $\qty{100\pm 0.3}{\MeV}$.  
A small fraction of particles that did not reach the final energy remained near \qty{18}{\MeV}.  
Figure~\ref{fig:FinalDist} shows the final distribution.  

If the rms-emittance of a coasting beam is defined as the standard deviation of the energy spread multiplied by the revolution time, its values at injection ($t = 0$), at the boundary energy ($t = T$), and the final energy ($t = 2T$) are 
\num{0.0114}, \num{0.0122}, and \qty{0.0159}{\eV\second}, respectively.  


\subsection{bucket characteristics and consistency with design}

\begin{figure}[thb]
  \centering
  \includegraphics[scale=1.0]{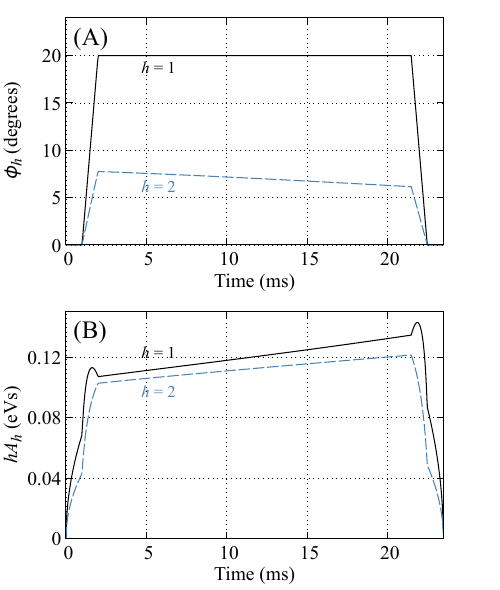}\\
  \caption{Evolutions of (A) synchronous phase and (B) $h \times$ bucket area for $h = 1$ and $h = 2$.  }
  \label{fig:AreaPhiS}
  \end{figure}

The synchronous phase and bucket areas for $h = 1$ and $h = 2$ were computed at each simulation time point using Eqs.~(\ref{eq:Harmonics}), (\ref{eq:DefineRh}), (\ref{eq:SinPh}), (\ref{eq:Area}), and (\ref{eq:BucketFormFactor}), as shown in Fig.~\ref{fig:AreaPhiS}.  
The bucket form factor $a(\phi)$ was evaluated via numerical integration; the approximations given in 
Eqs.~(\ref{eq:RhApproach}) and (\ref{eq:ApproxA}) were not used.  

As expected, the synchronous phase $\phi_2$ remains nearly constant, varying only slightly from \ang{7.7} to \ang{6.2}.  
Moreover, the longitudinal acceptances for the fundamental and second-harmonic energies evolve in close agreement, with the ratio $S_2 = 2A_2/A_1$ staying close to unity throughout the simulation.  
This consistency supports the design principles described in the previous section.  

The gradual increase in the fundamental bucket area shown in Fig.~\ref{fig:AreaPhiS} could, in principle, be mitigated by adjusting the rf voltage envelope, but this variation is small and does not significantly impact the simulation results.  

\section{Discussion \label{sec:Discussion} }

\subsection{Extension to multi-harmonic operation}

As mentioned in Sec.~\ref{sec:Introduction}, 
energy ranges corresponding to higher-order harmonics such as the fourth, eithth, and beyond are sequentially connected under the condition given by Eq.~(\ref{eq:feq2f}).
This suggests that particles can be accelerated from the bottom of a $2^n$-th harmonic energy range to the top of the fundamental energy range in $(n+1)$ rf-sweep cycles.

\begin{figure}[thb]
  \begin{center}
  \includegraphics[scale=1.0]{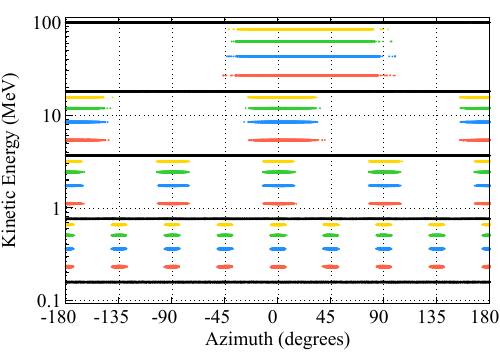}\\
  \caption{Longitudinal phase-space footprints of the beam, plotted at intervals of $T/5$, 
  during the multiharmonic acceleration from the eighth harmonic energy range to the fundamental. 
  Colors represent time slices in the order: black ($t=0$), red ($t=0.2T$), blue ($t=0.4T$), green ($t=0.6T$), 
  yellow ($t=0.8T$), and then repeat every interval of $T$. 
  The repeated color scheme reflects the periodic nature of the acceleration cycle, 
  and may result in multiple bunches at different stages appearing with the same color.}
  \label{fig:FootPrintH8}
  \end{center}
  \end{figure}

To assess the extended applicability of the scheme, an additional simulation was performed.
The rf pattern and machine parameters were kept unchanged, but the injection energy was reduced to \qty{159.3}{\keV}, which corresponds to the lower boundary of the eighth harmonic energy range (see Fig.~\ref{fig:SimEs8}).
Figure~\ref{fig:FootPrintH8} displays the beam's longitudinal phase-space evolution. 

The beam was first accelerated by the eighth harmonic rf buckets, followed by the fourth-, second-, and fundamental rf buckets in each subsequent interval of duration $T$.
After a total simulation time of $4T$, all macroparticles successfully reached the final energy of \qty{100}{MeV}.

These results confirm that the BPA scheme remains effective up to the eighth harmonic range.

\subsection{Above transition energy} \label{sec:Above}

\begin{figure}[thbp]
  \centering
  \includegraphics[scale=1.0]{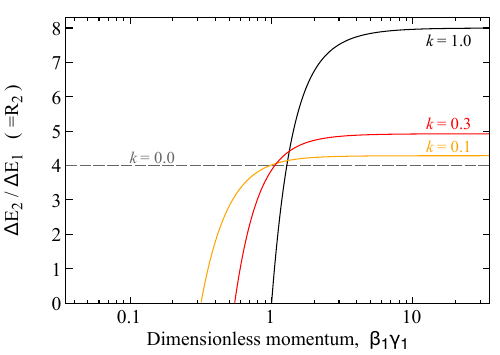}
  \caption{Ratios of one-turn energy gains $\Delta E_2/\Delta E_1$ as a function 
           of the dimensionless synchronous momentum $\beta\gamma$ above transition. 
           Different colors indicate the value of field index~$k$. 
  }
  \label{fig:PlotR2b}
  \end{figure}

Barber-pole acceleration is also applicable above transition energy.  
In this regime, the synchronous energy of higher harmonics becomes larger than that of the fundamental, and the acceleration ratio $R_2$ exceeds unity (see Fig.~\ref{fig:PlotR2b}).
In the ultrarelativistic regime, $R_2$ asymptotically approaches 
  \begin{equation*}
  R_h\ \simeq
  \ h^{k+2}  	\quad,\ \beta\gamma \gg 1\,.  
  \end{equation*}
which is the analogous to Eq.~(\ref{eq:RhApproach}) for the above-transition case.  

Above transition, the synchronous phase $\phi_2$ is greater than $\phi_1$, and it becomes natural to reverse the evaluation scheme used below transition: 
instead of comparing higher harmonic buckets against the fundamental ($h=1$), we now regard the higher harmonic (e.g., $h=2$) as the reference, and evaluate the fundamental bucket in relation to it.  
This change of reference automatically leads to the substitutions 
$R_h \mapsto R_h^{-1}$, $S_h \mapsto S_h^{-1}$, and $\sin\phi_1 \mapsto \sin\phi_h$ in the relevant expressions.

For instance, the synchronous phase relation from Eq.~\eqref{eq:SinPh} becomes
  \begin{equation*}
  \sin\phi_1 = R_2^{-1} \sin\phi_2,
  \end{equation*}
and the relative bucket area formula [Eq.~\eqref{eq:Sh}] is rewritten as
  \begin{equation*}
  S_h^{-1} = \frac{A_1}{hA_h} \sim \sqrt{R_h^{-1}} \,
  \frac{(1 - R_h^{-1} \sin\phi_h)(1 + \sin\phi_h)}{(1 + R_h^{-1} \sin\phi_h)(1 - \sin\phi_h)}.
  \end{equation*}

These relations can be interpreted using the same plot shown in Fig.~\ref{fig:RelativeArea} by applying the substitutions above and replacing $\phi_1 \mapsto \pi - \phi_h$ to reflect the fact that the accelerating phase lies within $[\frac{\pi}{2}, \pi]$ above transition.  

According to Fig.~\ref{fig:PlotR2b}, for $0 < k < 1$, $R_2$ lies in the range $4 < R_2 < 8$, or equivalently $0.125 < R_2^{-1} < 0.25$.  
In this range, Fig.~\ref{fig:RelativeArea} confirms that area matching ($S_2 \sim 1$) can still be achieved by choosing a synchronous phase around $\phi_2 \sim \ang{150}$.
 
\subsection{Effect of distributed rf cavities}

So far, the rf voltages $\overline{V}$ has been assumed to be common to both the fundamental and second-harmonic rf buckets, under the assumption that the rf cavity is localized at a single position in the ring.  
However, if the rf cavities are distributed around the ring, 
the phase relationship between them differs for the two harmonics, resulting in different effective rf voltage amplitudes experienced by the $h=1$ and $h=2$ buckets. 
Therefore, the rf cavities should, in principle, be localized at a single position in the ring.  

Conversely, this property can be utilized, if necessary, to introduce a controlled difference in the effective rf amplitudes for $h=1$ and $h=2$.   
Such an adjustment can be used to compensate for variations in $R_2$ or $S_2$.  
When $R_2$ varies significantly over the operational energy range, gradually varying the ratio of the effective rf amplitudes allows the synchronous phase $\phi_2$ to be kept constant.  
This adjustment can be realized by placing an auxiliary cavity with a small voltage amplitude $\overline{v}$ opposite to the main rf cavity.  
The two cavities operate in opposite phase for $h=1$ and in phase for $h=2$, so that the synchronous particles experience effective voltages 
  \begin{align*}
    \overline{V} - \overline{v} \quad & \text{for } h=1\,, \\
    \overline{V} + \overline{v} \quad & \text{for } h=2\,. 
  \end{align*}

\subsection{Collective effects}
When applying BPA to high-intensity beams, potential collective effects must be carefully considered~\cite{ChaoCollective,NgLecture}.  
In particular, since the BPA scheme is most effective in the energy region where $R_2$ remains nearly constant, it naturally targets relatively low beam energies, where space-charge effects become more significant.  
Moreover, because the scheme involves a debunching and a rebunching process, possible coasting-beam instabilities should also be taken into account.  
As an example of bunched-beam effects, we next consider possible resonant coupling between the betatron oscillations of different bunches.  

Barber-pole acceleration produces a heterogeneous beam structure, with bunches of distinctly different velocities circulating together in the ring.  
At any observation point where the beam encounters impedance, bunches arrive in a specific temporal sequence.  
An $h=2$ bunch (say, bunch 2A) arrives first, followed by the $h=1$ bunch after a short delay of $(\phi_1-\phi_2)/(2\pi\frf)$.  
Half an rf period later, the other $h=2$ bunch (2B) arrives, again followed by the $h=1$ bunch after the same short delay.  
This four-bunch sequence (2A$\rightarrow$1$\rightarrow$2B$\rightarrow$1) repeats every rf cycle.  
The bunches are centered at different closed-orbit radii, corresponding to their respective synchronous energies, and therefore pass through the observation point at slightly different transverse positions.  

The wake field generated by the two $h=2$ bunches produces spectral components at 
  \begin{equation*}
  \frf\left(n\pm\frac{\mu+Q}{h}\right)\,, 
  \end{equation*}
where $h=2$ and $\mu=0,1$ denote the harmonic number and the coupled-bunch mode number, respectively.  
Fortunately, these frequencies do not coincide with the betatron sidebands of the $h=1$ beam and therefore do not resonantly excite its transverse oscillations.  

Nevertheless, the BPA scheme involves highly nontrivial longitudinal and transverse beam dynamics, and the present discussion covers only limited set of possible effects.  
A more comprehensive evaluation of collective phenomena will be required for detailed design studies of high-intensity applications.

\section{Conclusion \label{sec:Conclusion} }

We have proposed a novel acceleration scheme, termed \textit{barber-pole acceleration} (BPA), in which beams are accelerated across the first- and second-harmonic rf buckets through a continuous frequency sweep.  
A key advantage of this method is that, by setting the rf frequency sweep range to a factor of two, 
one can accelerate beams over an energy range corresponding to a fourfold change in revolution frequency.  
In practice, the actual beam acceleration range need not exactly match this full span if injection into or extraction from moving buckets are possible.  
A twofold rf bandwidth is technically feasible and has already been realized in many existing accelerator systems~\cite{HGC,COSY,KEKFFA,LEIR}.  

The concept can be extended to schemes that deliberately skip the fundamental harmonic.  
For example, 
by choosing an rf frequency span of a factor of 1.5 instead of 2, one may connect harmonic chains such as $h=2\leftrightarrow 3$ and $h=4\leftrightarrow 6\leftrightarrow 9$.  
This approach may be advantageous when minimizing the required frequency span is of particular importance, although its practical merits depend on specific design trade-offs.  

As shown in Sec.~\ref{sec:Above}, the scheme can also operate above transition.  
The required orbit excursion, however, differs substantially between the two regimes.  
In BPA, the beam revolution frequency must change by at least a factor of two, ideally four.  
Under this condition, the ratio of orbit radii is approximately $r_1/r_2\sim 2^{1/k}$ below transition, whereas it becomes roughly $r_2/r_1\sim 2$ above transition.  
Thus, while the method remains valid in principle, its application above transition is likely less attractive from an engineering and cost perspective.  

Omitting the debunch-rebunch process altogether may also be considered.  
In this approach, the beam is transferred directly from one bucket to the next, which avoids potential coasting-beam instabilities and further reduces the overall acceleration time.  
A drawback, however, is that one of the two $h=2$ buckets cannot be utilized in this scheme.  
Depending on the specific situation, this approach can be adopted as a viable alternative.  

Experimental verification of the proposed scheme using an existing FFA machine is currently under consideration, aiming to assess its practical feasibility and operational performance.


\appendix
\section{Editing rf pattern of a scaling FFA \label{sec:PatternEdit} }

In a scaling FFA, the magnetic field $B$ is proportional to $r^k$, where $r$ is the radius from the machine center and $k$ is a constant called the field index. Under this condition, the closed orbits at different energies are geometrically similar.  
The momentum $p$ and the average orbit radius $r$ are related by
  \begin{equation*}
  \frac{r}{\refv{r}} = \left( \frac{p}{\refv{p}} \right)^\alpha\,,  
  \end{equation*}
where $\alpha = 1/(k+1)$ is the momentum compaction factor, and the subscript “ref” denotes the reference value used for normalization.  
The revolution frequency $f$ of a particle on the closed orbit is then given by
  \begin{equation}
  \frac{f}{f_\mathrm{ref}} = \left( \frac{E_\mathrm{ref}}{E} \right)
                             \left( \frac{p}{p_\mathrm{ref}} \right)^{1-\alpha} \,. 
  \label{eq:AppFofE}
  \end{equation}
The orbital angular momentum, $L = r p$, evolves as
  \begin{equation*}
  \frac{L}{L_\mathrm{ref}} = \left( \frac{p}{p_\mathrm{ref}} \right)^{1+\alpha}\,.
  \end{equation*}
Differentiating the logarithm of the above expression yields
  \begin{equation*}
  \frac{\mathrm{d}L}{L} = (1 + \alpha)\frac{\mathrm{d}p}{p}
  = (1 + \alpha) \frac{E\, \mathrm{d}E}{c^2 p^2}\,.
  \end{equation*}
Returning to the relation $L = r p$ and using $f = v / (2\pi r)$, the result is simplified to
  \begin{equation*}
  \mathrm{d}L = \frac{1 + \alpha}{2\pi f} \, \mathrm{d}E\,.
  \end{equation*}

If the synchronous energy gain per turn is defined as
  \begin{equation*}
  \Delta E = \frac{1}{f_\mathrm{syn}} \frac{\mathrm{d}E_\mathrm{syn}}{\mathrm{d}t}\,, 
  \end{equation*}
then the synchronous angular momentum evolves over time as
  \begin{equation*}
  \mathrm{d}L_\mathrm{syn} = \frac{1 + \alpha}{2\pi} \Delta E \, \del{t}\,,
  \end{equation*}
and hence,
  \begin{equation}
  L_\mathrm{syn}(t) = L_\mathrm{syn}(0) + \frac{1 + \alpha}{2\pi} \int_0^t \Delta E(t)\, \mathrm{d}t\,.  
  \label{eq:LofE}
  \end{equation}
By solving the above expression, the synchronous momentum and the corresponding rf frequency can be obtained as functions of time.

As an example, suppose that $\overline{V} = v_0 + \dot{v} t$ and $\phi_\mathrm{s} = \phi_0 + \dot{\phi} t$, where $v_0$, $\dot{v}$, $\phi_0$, and $\dot{\phi}$ are constants with $\dot{\phi} \neq 0$, then:
  \begin{equation}
  \int_0^t \Delta E(t)\, \mathrm{d}t 
  = \, \mathrm{Im} \left[ 
    \left( \frac{\dot{v} - i\dot{\phi} \overline{V}}{\dot{\phi}^2} \right)
    e^{i\phi_\mathrm{s}} 
  \right]_0^t\,.
  \label{eq:FreqPattern}
  \end{equation}

\section{Tracking algorithm}\label{sec:Algorithm}

This section describes the simulation procedure used for barber-pole acceleration.  
Since particles may occupy rf buckets with different harmonic numbers, special care is required in defining the tracking variables.  
Here, we adopt a method that updates each particle's azimuthal position $\psi_i$ and energy $E_i$ on an rf-cycle-by-cycle basis.  
The azimuthal coordinate $\psi_i$ is defined as the angular distance remaining before the particle reaches the rf cavity.  
The simulation assumes a single rf cavity in the ring and neglect transverse motion.  

Let the rf frequency and voltage amplitude during a given rf cycle be $\frf$ and $\overline{V}$, respectively.  
For each macro-particle, the following steps are repeated to update its coordinates after one rf cycle: 
  \begin{enumerate}
  \item Initialize the remaining rf phase, $\delta\phi=2\pi$.  
  \item Calculate the slippage coefficient $\eta'=f(E_i)/\frf$.  
  \item Compute the azimuthal distance the particle travels until the next rf zero-crossing: \\
      	$\Delta\psi=\eta'\delta\phi$.  
  \item if $\Delta\psi \le \psi_i$, update the coordinates as: \\$\psi\leftarrow \psi_i-\Delta\psi$, exit loop.  
  \item Otherwise, the particle reaches the rf cavity within this rf cycle: \\
     	$\delta\phi\leftarrow \delta\phi-\psi_i/\eta'$, $E_i\leftarrow E_i+q\overline{V}\sin(2\pi-\delta\phi)$, 
     	$\psi_i\leftarrow 2\pi$.  Return to step (2).  
  \end{enumerate}
This procedure is applied to all macro-particles, and then the rf frequency and voltage are updated before proceeding to the next rf cycle.

\bibliography{BarberPole}

\end{document}